\documentclass{ws-p8-50x6-00}

\begin{document}

\title{Particle Correlations at LEP}

\author{T.~H. Kress}

\address{University of California, Riverside and OPAL Collaboration \\
E-mail: Thomas.Kress@cern.ch}

\maketitle

\abstracts{
Particle correlations are extensively studied to obtain information about the
dynamics of hadron production. 
From 1989 to 2000 the four LEP collaborations recorded more than 16 million 
hadronic $\mathrm{Z^{0}}$ decays and several thousand $\mathrm{W^{+}W^{-}}$
events.
In $\mathrm{Z^{0}}$ decays, two--particle correlations were analysed in detail 
to study Bose--Einstein and Fermi--Dirac correlations for various particle 
species. 
In fully--hadronic $\mathrm{W^{+}W^{-}}$ decays, particle correlations were
used to study whether the two W bosons decay independently.
A review of selected results is presented.}

\section{Introduction}
The analysis of particle correlations in high energy interactions gives
important information about the hadron production mechanism, complementary
to studies of global event properties and single--particle distributions.

The LEP $\mathrm{e^{+}e^{-}}$ collider provides an ideal environment for such 
studies.
From 1989 until 1995, LEP operated at centre--of--mass energies around 91 GeV
which allowed each of the four experiments, ALEPH, DELPHI, L3 and OPAL, to 
record more than four million hadronic $\mathrm{Z^{0}}$ decays.
From 1996, after the collider energy had been increased above the WW threshold,
until the end of LEP four years later, each experiment recorded 
about ten thousand $\mathrm{W^{+}W^{-}}$ events.
A hadronic decay of a Z or W boson leads to some dozen particles in the
final state, mostly charged pions and photons from the decay of the 
$\mathrm{\pi^{0}}$ mesons, but also, to a lesser extent, to kaons, protons and 
lambda hyperons, which allows to study particle correlations in detail.

Bose--Einstein correlations (BEC) between identical bosons are a well 
established phenomenon in high energy physics experiments and are often 
considered to be an equivalent of the Hanbury Brown \& Twiss\cite{bib:hbt} 
(HBT) effect in astronomy describing the interference of photons emitted 
incoherently. 
An alternative approach was proposed by Andersson et.~al.\cite{bib:lund}
taking into acount the dynamics of hadron formation in a coherent production
process within the framework of the Lund string model related to the
symmetrisation of the quantum--mechanical amplitude.

Bose--Einstein correlations lead to an enhanced production of identical 
boson pairs with a small four--momentum difference 
$\mathrm{Q^{2}=-(p_{1}^{\mu}-p_{2}^{\mu})^{2}}$.
Traditionally it is studied using a two--particle correlation function
$\mathrm{C(p_{1},p_{2}) = \rho_{2}(p_{1},p_{2})/\tilde{\rho}_{2}(p_{1},p_{2})}$, 
where $\mathrm{\rho_{2}}$ and $\mathrm{\tilde{\rho}_{2}}$ are the two--particle 
densities with and without BEC, respectively.
For the construction of the reference sample $\mathrm{\tilde{\rho}_{2}}$ 
frequently a MC model without BEC is used.   
Following the pioneering analysis of Goldhaber, Goldhaber, Lee and 
Pais\cite{bib:gglp} (GGLP), a correlation 
function of type $\mathrm{C(Q) = 1 + \lambda\exp(-Q^{2}r^{2})}$ is often used 
to yield a value for r, which is interpreted to be the emitter radius. 
The factor $\mathrm{\lambda}$ measures the strength of the BEC effect but 
sometimes absorbs experimental inpurities.
Extra terms in the parametrisation are occasionally used to account
for imperfections in the description of other correlations in
the reference sample.

\section{Particle Correlations in $\mathbf{Z^{0}}$ Decays}
\subsection{Bose--Einstein Correlations in Pion Pairs}
\begin{figure}[h]
\epsfxsize=14.0pc \epsfysize=12.0pc \epsfbox{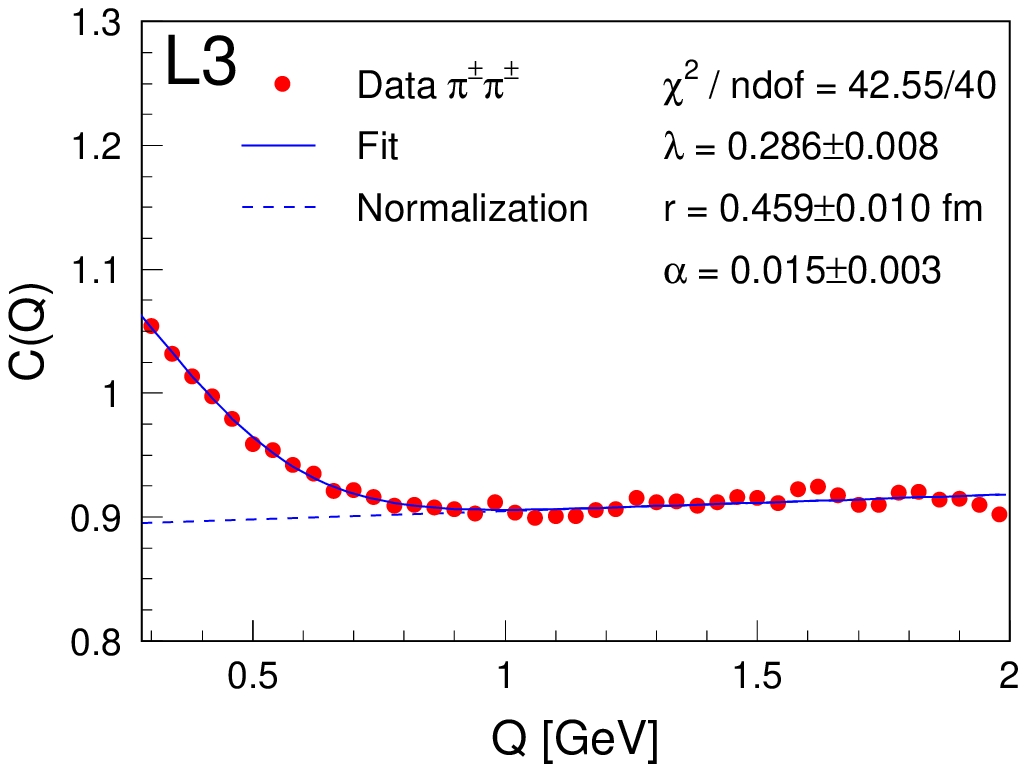}
\epsfxsize=14.0pc \epsfysize=12.0pc \epsfbox{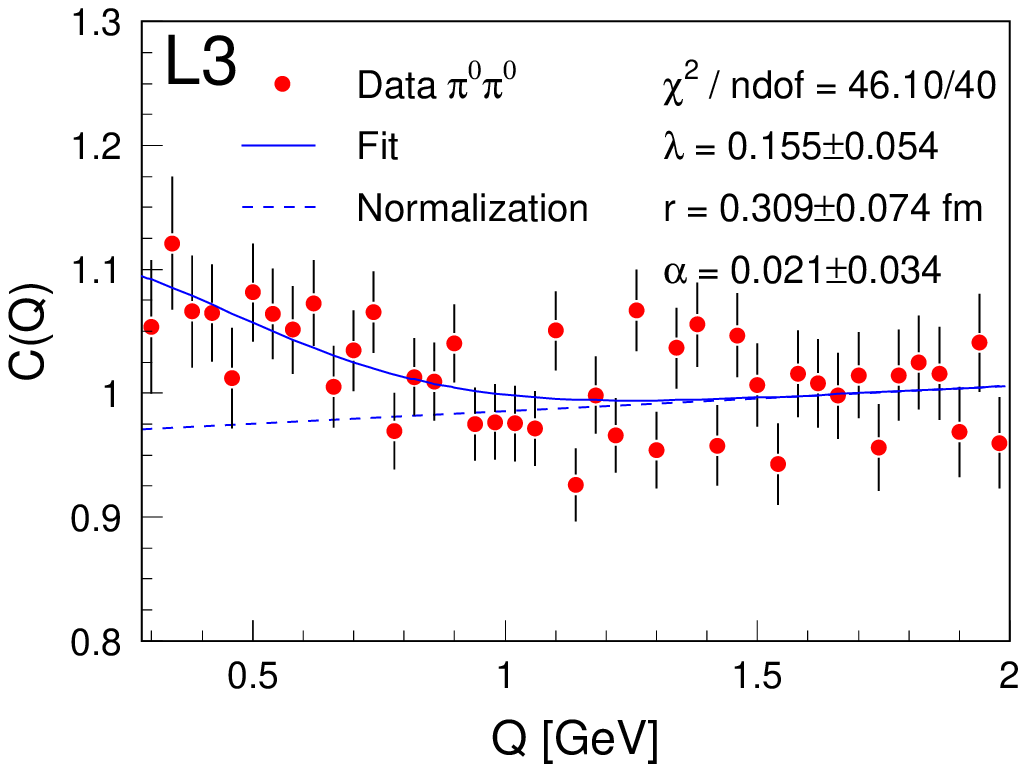}
\caption{Correlation function C(Q) for pairs of charged and neutral 
pions.\label{fig:pion}}
\end{figure}

Fig.~\ref{fig:pion} shows recent L3 measurements\cite{bib:l3_pion} of the correlation
function for charged (left) and neutral (right) pion pairs using a MC without BEC as
the reference samples.
For both the charged and neutral pion pairs an enhancement at low Q is 
clearly visible. 
Using a parametrisation \`{a} la GGLP, the obtained source radius 
for neutral pions is tending to be smaller than r for
charged pions, as qualitatively expected in the Lund string
model\cite{bib:zalewski}.

\subsection{Elongation of the Pion Source}

A difference in the Bose--Einstein correlation length longitudinally and
transversely with respect to the jet axis in $\mathrm{e^{+}e^{-}}$ 
annihilation, arises naturally in a model for Bose--Einstein correlations 
based on the Lund string model\cite{bib:lund_2d}.

\begin{figure}[h]
\begin{center}
\epsfxsize=14.0pc \epsfysize=13.pc \epsfbox{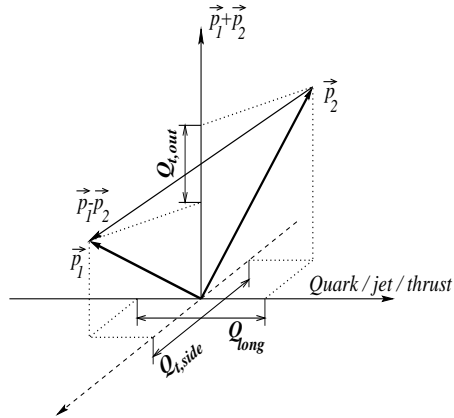}
\end{center}
\caption{The Longitudial Centre--of--Mass System.\label{fig:lcms}}
\end{figure}

In the Longitudinal Centre--of--Mass System, which is defined for each
pair of particles as the system in which the sum of the two particles'
momenta is perpendicular to the thrust axis of the process, the
three--momentum difference $\mathrm{\vec{Q}}$ is decomposed in three
components as illustrated in Fig.~\ref{fig:lcms}.
An analysis with a GGPL parametrisation, separately for the components 
$\mathrm{Q_{long}}$, $\mathrm{Q_{t,side}}$ and $\mathrm{Q_{t,out}}$, gives 
access to the transverse and longitudinal emitter radius.
In spite of different methods for the construction of the reference sample,
the results\cite{bib:lep_2d} of DELPHI, L3 and OPAL consistently demonstrate 
that the particle emission source is elongated along the direction of motion
with a ratio of transverse to longitudinal radius in the range 0.6---0.8.

\subsection{Bose--Einstein Correlations in Kaon Pairs}
\begin{figure}[h]
\epsfxsize=14.0pc \epsfysize=12.0pc \epsfbox{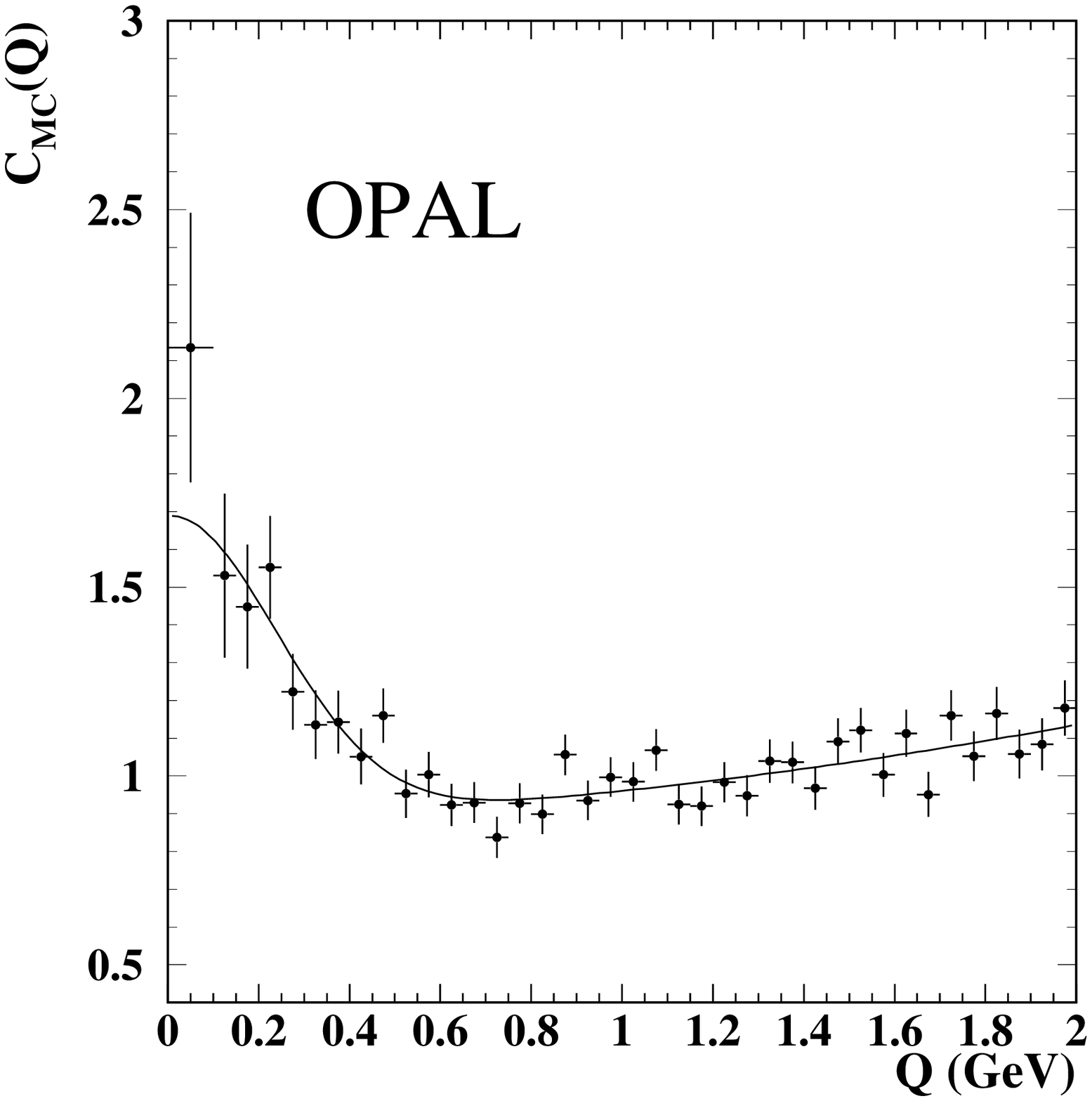}
\epsfxsize=14.0pc \epsfysize=12.0pc \epsfbox{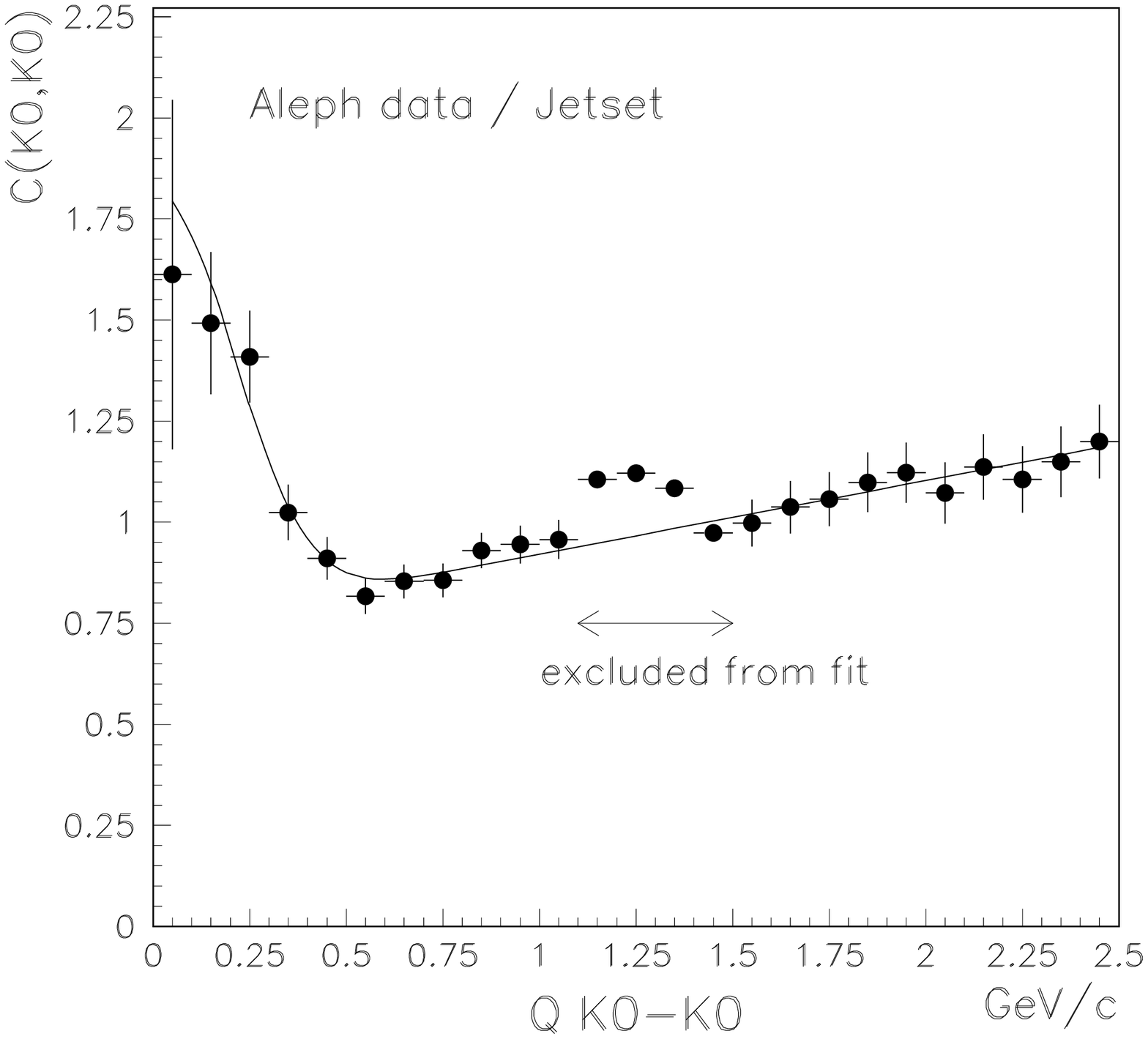}
\caption{Correlation function C(Q) for charged kaon pairs (left)
and $K_{S}$ pairs (right). Both analyses use a MC without BEC to
construct the reference sample.\label{fig:kaon}}
\end{figure}

Bose--Einstein correlations have also been established in charged and
neutral kaon pairs at LEP.
The left--hand side of Fig.~\ref{fig:kaon} shows the correlation function
of selected pairs of charged kaons obtained by OPAL\cite{bib:opal_kaon}.
The kaons were identified using information of the specific ionisation
energy loss (dE/dx) in the large volume jet chamber.
For neutral kaons an enhanced production at low Q is 
expected\cite{bib:lipkin} if the C parity of the system is +1 as 
for two $\mathrm{K_{S}}$ mesons.
On the right--hand side of Fig.~\ref{fig:kaon} the correlation function for
pairs of $\mathrm{K_{S}}$ is shown as obtained by ALEPH\cite{bib:aleph_kaon}.
The $\mathrm{K_{S}}$ mesons were identified by using the 
$\mathrm{M_{\pi^{+}\pi^{-}}}$ invariant mass spectrum for candidates with a 
secondary vertex.

\subsection{Fermi--Dirac Correlations}
\begin{figure}[h]
\epsfxsize=14.0pc \epsfysize=15.0pc \epsfbox{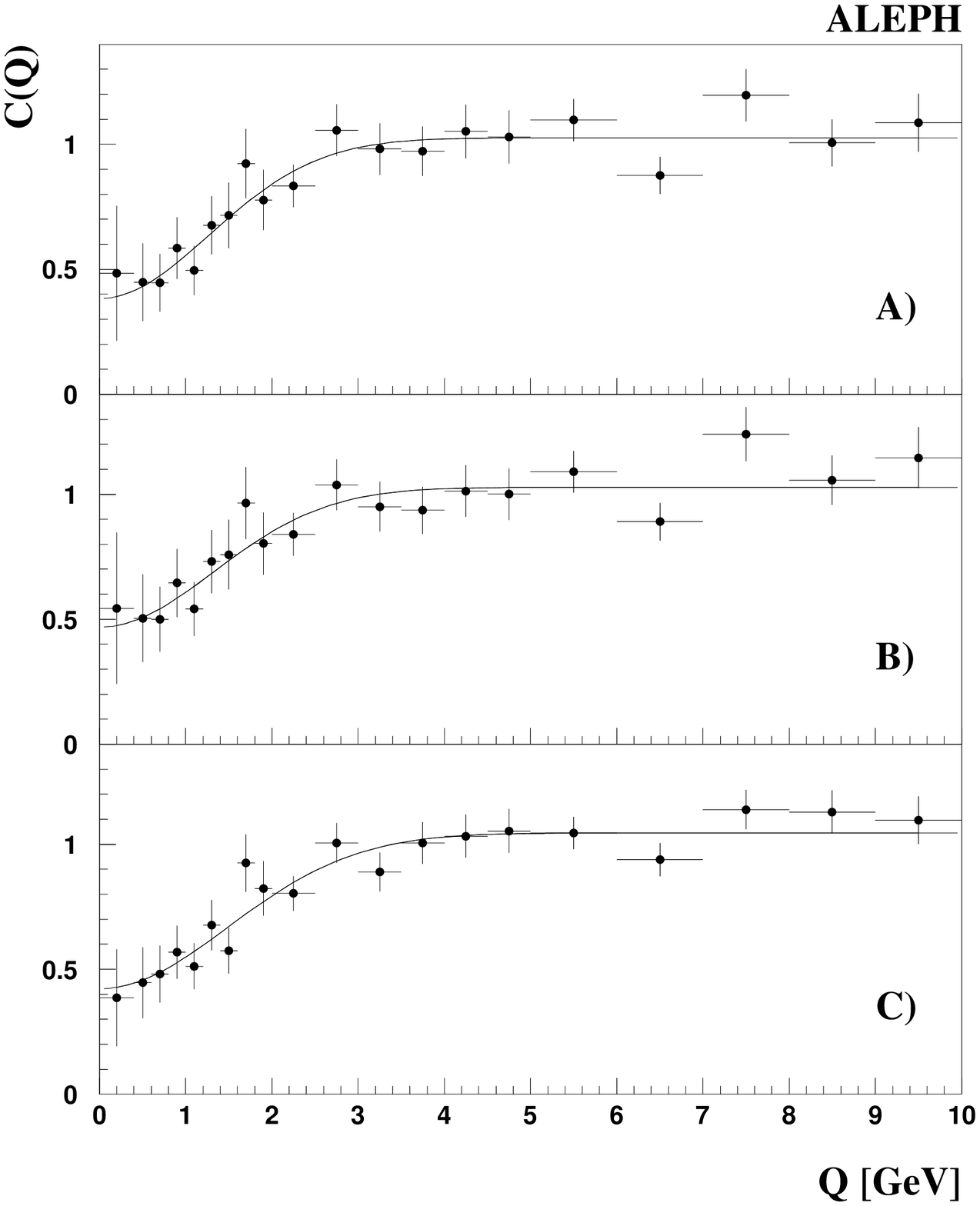}
\epsfxsize=14.0pc \epsfysize=14.7pc \epsfbox{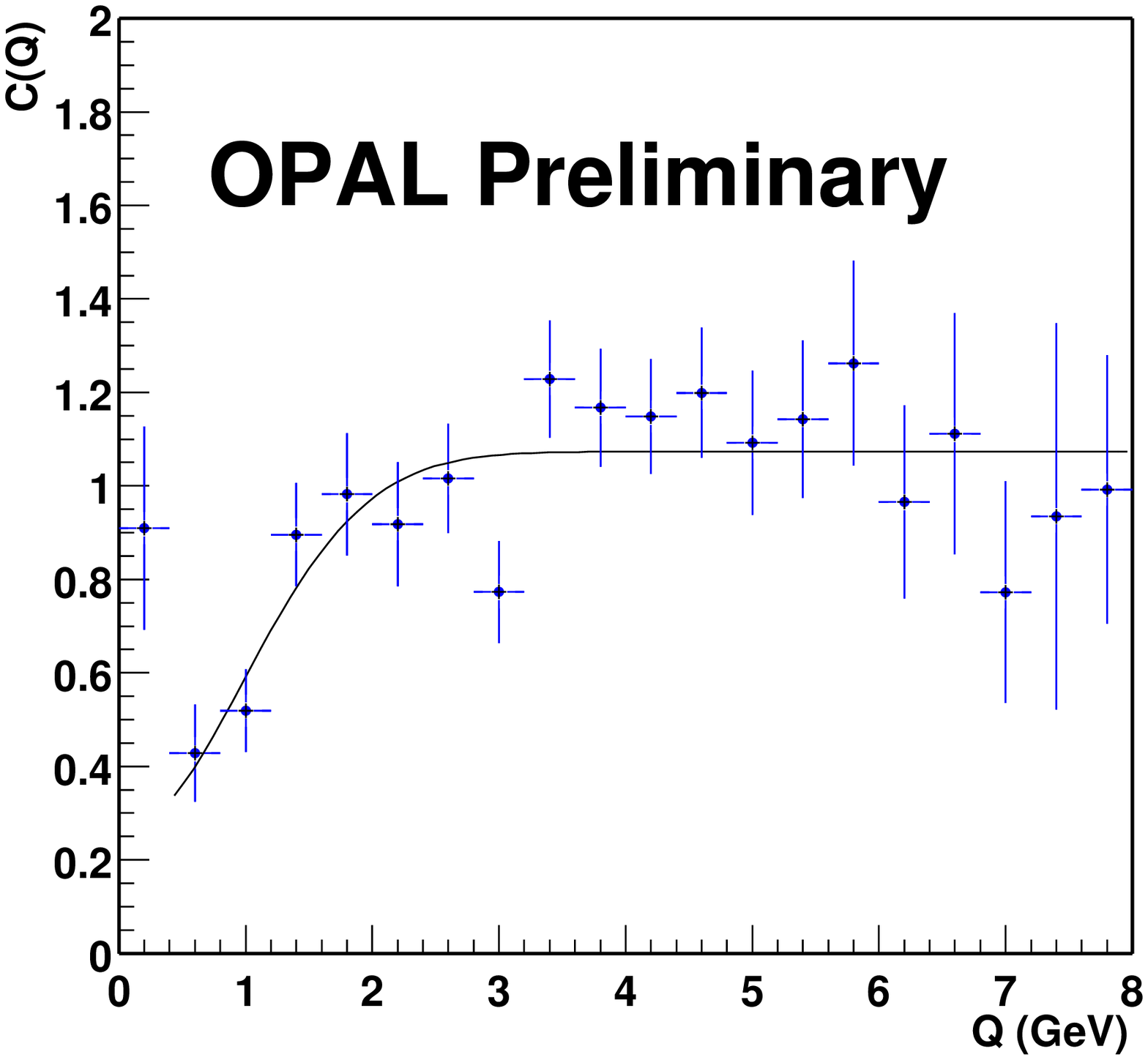}
\caption{Correlation function for $\Lambda^{0}$, $\bar{\Lambda}^{0}$
(left) and anti--proton (right) pairs.\label{fig:baryon}}
\end{figure}

Recently it has been proposed\cite{bib:alexander_fd} to extract an
emitter dimension for pairs of equal baryons by utilising the
Fermi--Dirac exclusion principle. 
The correlation function can be parametrised by an equation similar 
to the GGLP parametrisation with the plus sign replaced by a minus 
sign.
Antisymmetrising the total wave function yields four states, three 
of which are antisymmetric in space and symmetric in spin. Thus, for
an incoherent source, C(Q) should decrease to a value 1/2 in the 
limit as $\mathrm{Q \rightarrow 0}$.
The left--hand side of Fig.~\ref{fig:baryon} shows the ALEPH 
results\cite{bib:aleph_lambda} for the correlation function from lambda and 
anti--lambda pairs. The hyperons were identified with a similar method
as described for the $\mathrm{K_{S}}$ mesons.
Using three different methods for the construction of the reference
sample (A, B, C), a depletion at low Q is observed in all cases.
As shown on the right--hand side of Fig.~\ref{fig:baryon}, a depletion 
for baryons at low Q is confirmed by preliminary results from
the OPAL collaboration\cite{bib:opal_proton} using pairs of anti--protons, 
identified by dE/dx information.

\subsection{Mass Dependence of the Emitter Radius}
\begin{figure}[h]
\epsfxsize=14.0pc \epsfysize=12.0pc \epsfbox{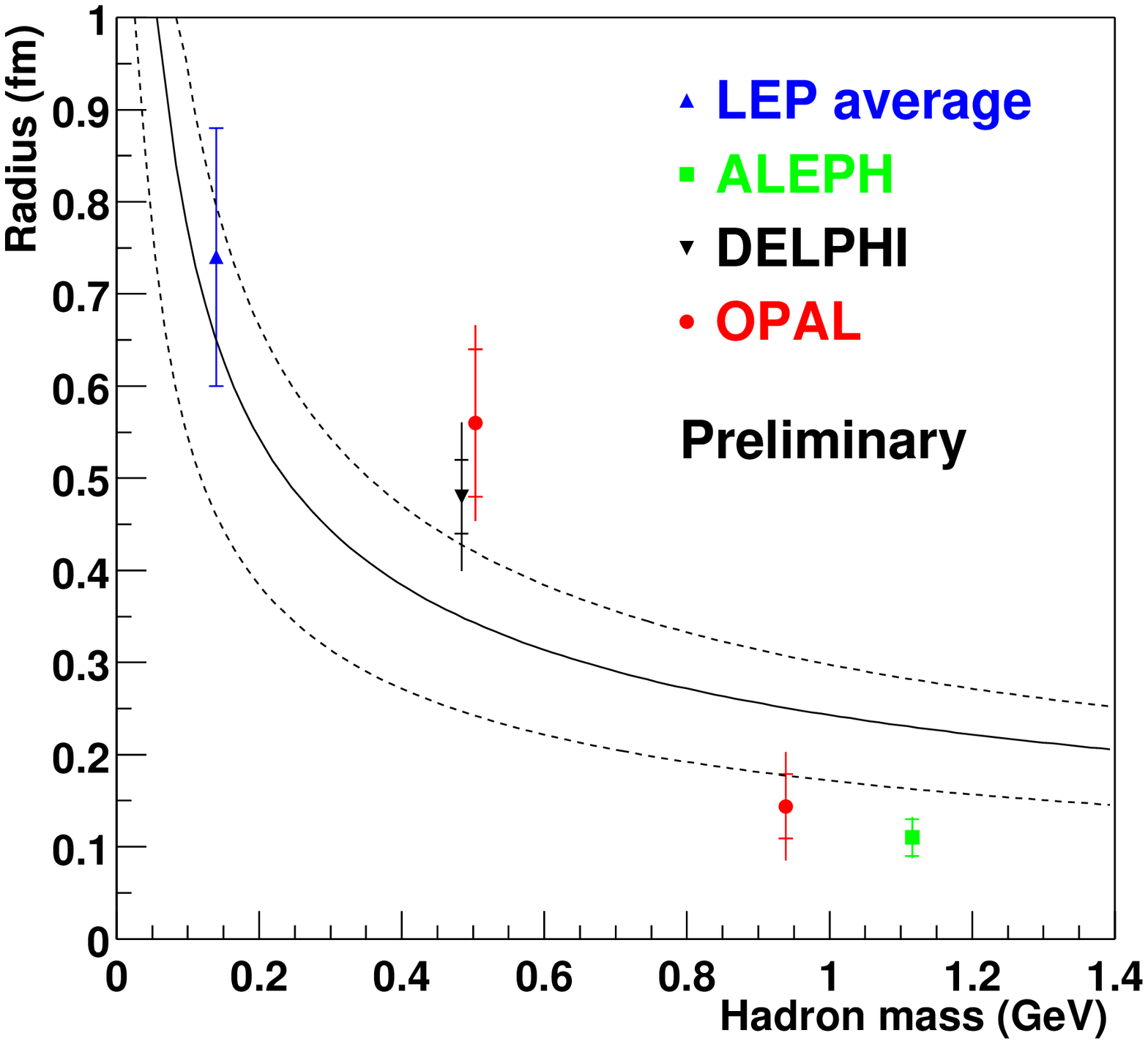}
\epsfxsize=14.0pc \epsfysize=12.0pc \epsfbox{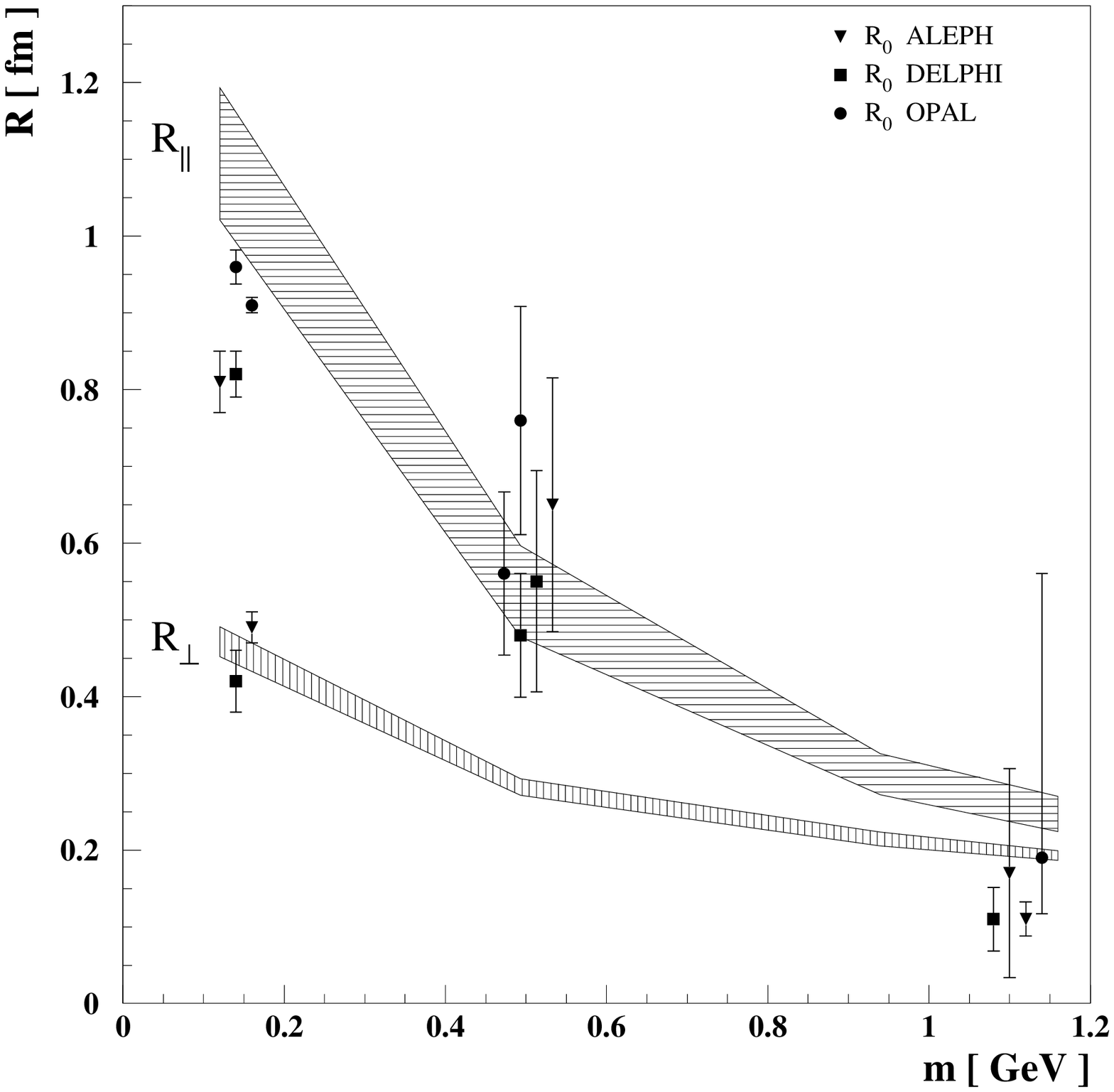}
\caption{Measured source radii as a function of the hadron mass
         compared to predictions based on the Heisenberg uncertainty 
         relations (left) and on the inside--outside cascade model 
         (right) as explained in the text.\label{fig:Rmass}}
\end{figure}

On the left--hand side of Fig.~\ref{fig:Rmass}, the measured source radii 
for pairs of pions (averaged by Alexander et.~al.\cite{bib:alexander_Rmass}), kaons,
lambda hyperons and (anti--)protons are plotted as a function of the hadron mass.  
A mass hierachy $\mathrm{r_{\pi} > r_{K} > r_{p,\Lambda}}$ is visible,
although the statement is only justified because of the baryon measurements.
The Lund string model in its basic form expects $\mathrm{r(m)}$ to
increase with m, thus the very small value obtained for $\mathrm{r_{\Lambda}}$
and $\mathrm{r_{p}}$ poses a challenge to this model\cite{bib:andersson_moriond}.
It has been shown\cite{bib:alexander_Rmass} that by applying the
Heisenberg uncertainty principles one can derive an expression for
$\mathrm{r(m)}$ that decreases with the hadron mass m, namely
\begin{math} r(m) = c\sqrt{\hbar\Delta t}/\sqrt{m}. \end{math}
In Fig.~\ref{fig:Rmass} the prediction is shown with $\mathrm{\Delta}$t
set to (0.5/1.0/1.5)$\cdot$10$^{-24}$ sec (lower dashed/solid/upper dashed
line), to represent a typical scale for strong interactions.
The prediction is able to qualitatively describe the data.
As pointed out by Alexander\cite{bib:alexander_moriond}, an
enormous energy density of the order of 10--100 GeV/fm$^{3}$ arises for
the baryons if the measured r in fact represents the emitter radius. 
Another interpretation\cite{bib:bialas_Rmass} of the data in the 
framework of the inside--outside cascade model avoids the potential
problem with the extrem energy density. 
A proportionality between the four--momentum of a produced particle
and the four--vector describing its space--time position at the 
freeze--out is commonly accepted in the description of high--energy
collisions. Provided that all particles are emitted from a tube of 
$\approx$ 1 fm in diameter at a constant proper time of $\approx$ 1.5 fm, 
the model can explain the data as shown in the right--hand side of
Fig.~\ref{fig:Rmass}.
In this approach the measured $\mathrm{r(m)}$ dependence is solely a
consequence of the strong correlation between $\mathrm{x^{\mu}}$
and $\mathrm{p^{\mu}}$.

\vspace*{1mm}

\section{Inter--WW Bose--Einstein Correlations}
\begin{figure}[h]
\epsfxsize=13.0pc \epsfysize=12.0pc \epsfbox{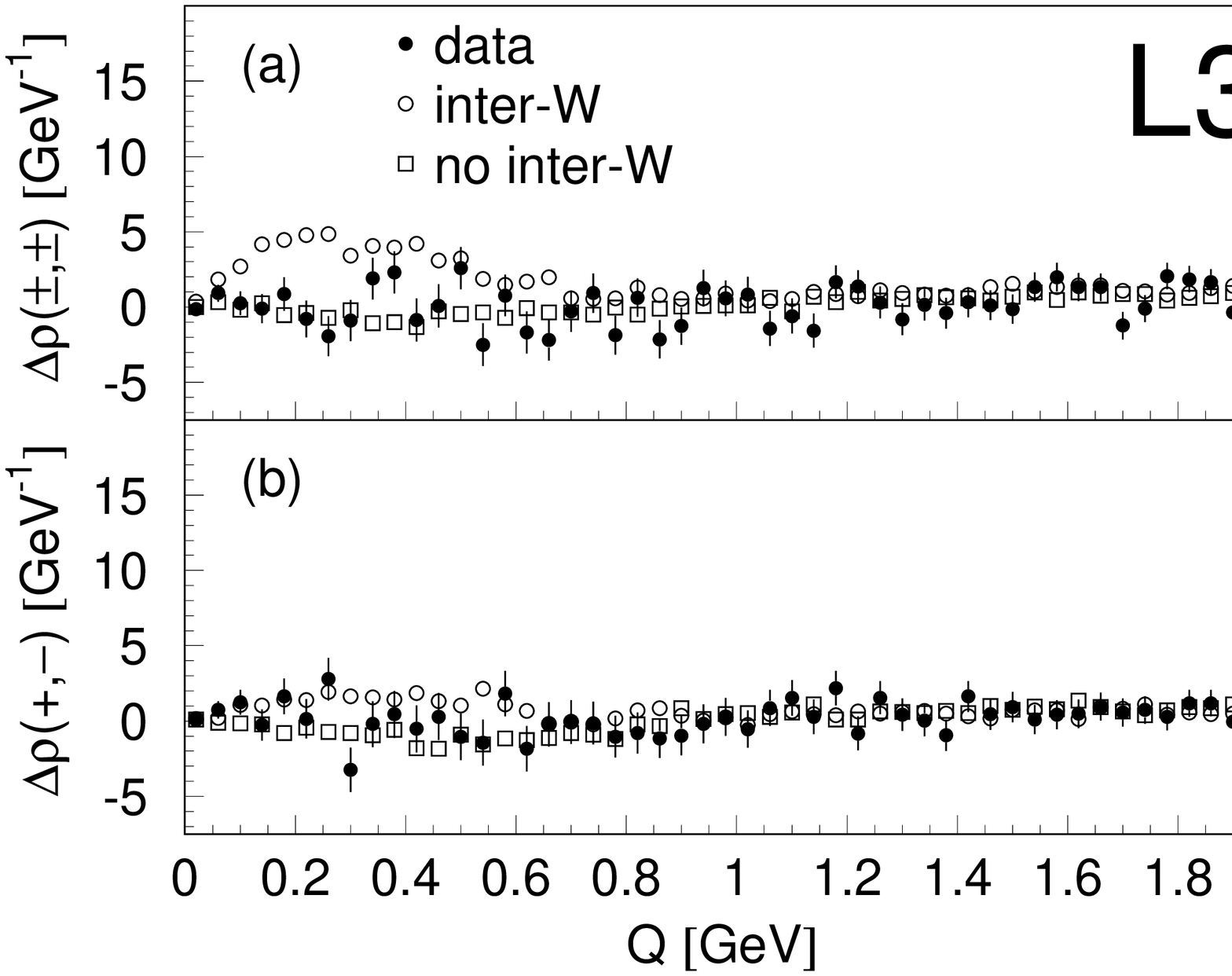}
\hspace*{2.0pc}
\epsfxsize=13.0pc \epsfysize=12.0pc \epsfbox{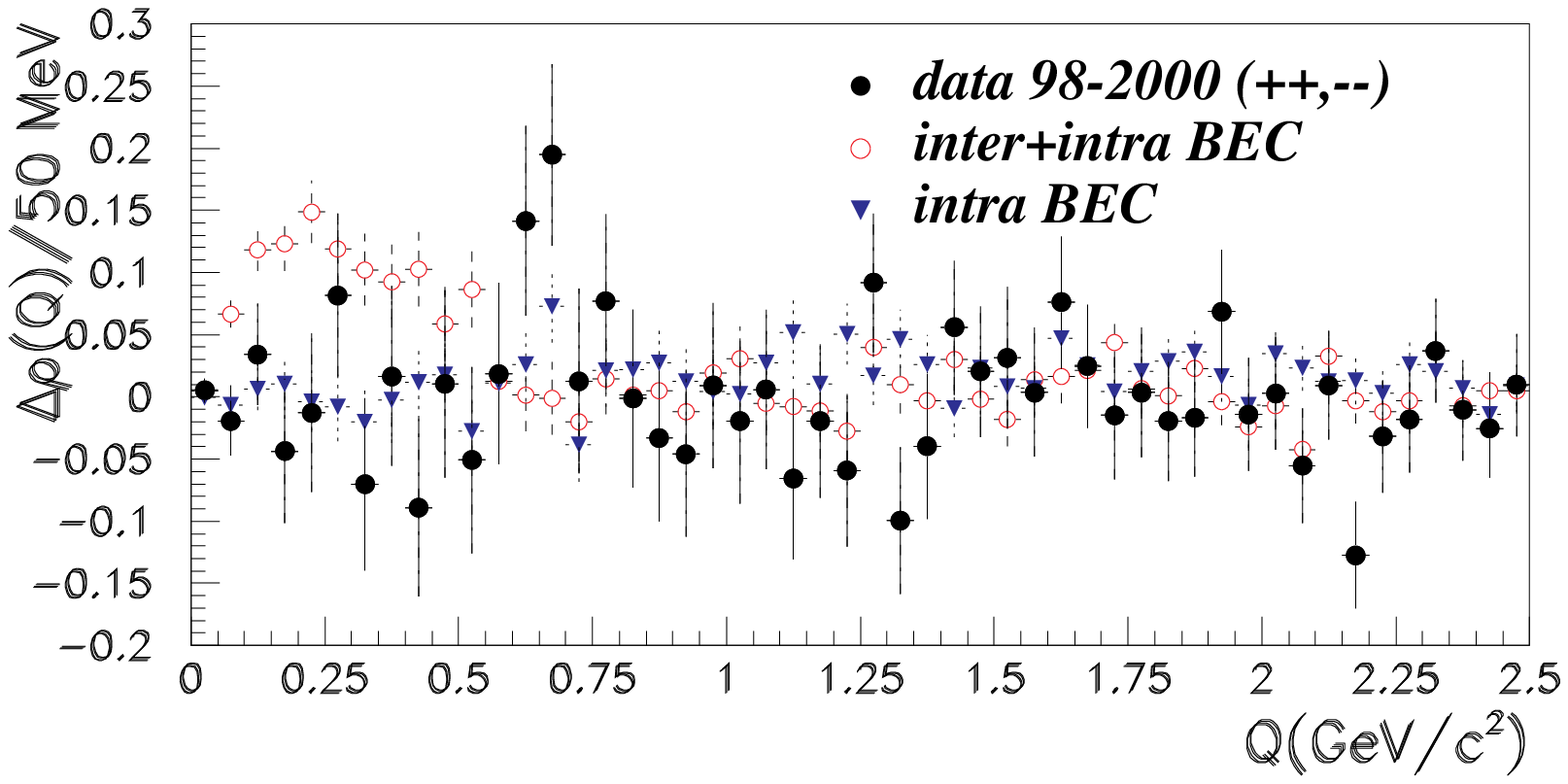}
\caption{Preliminary $\Delta\rho$ distributions from L3 (left) and DELPHI (right) 
compared to MC models with and without inter--WW BEC.\label{fig:Drho}}
\end{figure}

In $\mathrm{W^{+}W^{-} \rightarrow q\bar{q}q\bar{q}}$ events at LEP,
the products of the W decays in general have a significant 
space--time overlap as the separation of the their decay vertices
is small compared to characteristic hadronic distance scales.
The W boson mass, a fundamental parameter in the Standard Model, 
is determined from the corresponding jet masses and could
potentially be biased if Bose--Einstein correlations between the decay
products of the two W bosons exist. 
A robust framework to test the presence of such inter--WW BEC
was proposed in\cite{bib:cdwk}. If the $\mathrm{W^{-}}$ and 
$\mathrm{W^{+}}$ decay independently, then $\mathrm{\Delta\rho}$(Q)$\equiv$0
for all Q with the test distribution $\mathrm{\Delta\rho}$ defined as:
\begin{displaymath}
\Delta\rho = 
\rho^{WW\rightarrow 4q} - 2\cdot \rho^{W\rightarrow2q} - \rho^{WW_{mix}},
\end{displaymath}
with the two--particle densities $\mathrm{\rho^{WW\rightarrow 4q}}$
determined by the 
$\mathrm{W^{+}W^{-} \rightarrow q\bar{q}q\bar{q}}$ sample, 
$\mathrm{\rho^{W\rightarrow2q}}$ by the hadronic part of
semileptonic $\mathrm{W^{+}W^{-} \rightarrow q\bar{q} l\bar{\nu}}$ events 
and $\mathrm{\rho^{WW_{mix}}}$
from events build from two independent semileptonic events without
the leptonic parts and combining only particles originating from
different W's.
In Fig.~\ref{fig:Drho}, the L3 collaboration\cite{bib:l3_ww} compares the 
$\mathrm{\Delta\rho}$ distribution obtained from the data with two scenarios of the 
PYTHIA/PYBOEI\cite{bib:pyboei} Monte Carlo model with BEC. In the upper
plot the inter--WW BEC in the Monte Carlo
model can be seen as an enhancement of like--sign pairs in the low Q 
region. Unlike--sign pairs (lower plot) are artificially effected by
the technical implementation of the inter--WW BEC in PYBOEI. 
The data are consistent with no inter--WW correlations and the 
Monte Carlo which describes BEC between particles from different
W bosons in the same way as the correlations within the same W is
strongly disfavoured.
The same is true for similar results from DELPHI\cite{bib:delphi_ww}
as shown on the right--hand side of Fig.~\ref{fig:Drho}.
Both results are preliminary. 

\begin{figure}[h]
\begin{center}
\epsfxsize=13.0pc \epsfysize=12.0pc \epsfbox{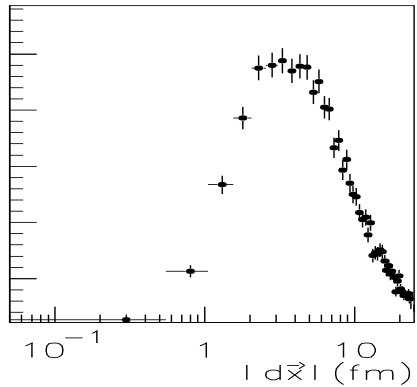}
\end{center}
\caption{The space--time distance of production vertices of direct equally charged
pions for pairs from different W's at $\sqrt{s}$=172 GeV.\label{fig:separation}}
\end{figure}

In the Lund model, Bose--Einstein correlations arise when identical 
bosons are produced close to each other within the same string.
Because of the strong correlation of production space--time and
momentum of the hadron in the inside--outside cascade, the
measured r is interpreted as the distance in the string where
the momentum spectra of the particles still overlap. 
In the absence of colour reconnection effects, particles from 
different W bosons are not produced in the same string.  
However, in addition to the coherent correlations inside a string a 
second correlation effect of an incoherent HBT type could be present.
An analysis\cite{bib:todorova} of the hadron formation 
within the Lund model shows that the
space--time distance of the production vertices for pairs of particles 
from different strings is of the order of several fm as illustrated in 
Fig.~\ref{fig:separation}.  
For such large distances any remaining inter--WW BEC effect would manifest
itself only at very low Q values\cite{bib:dewolf} which are hard to exploit
with the limited statistics of WW events at LEP.

\section{Summary}
More than 40 years after BEC entered the high energy physics stage, we now
have come to a better understanding of the effect.
It has become clear that one has to go beyond a naive HBT interpretation
by taking into account the dynamics of hadron production.
Since no firm theory exists to describe the hadronisation phase, we rely
on phenomenological models. To test such models, particle correlations provide 
us with details complementary to those obtained from global event properties 
and single--particle distributions.


%
\end{document}